\documentclass{article}
\usepackage{times,geometry,enumerate}
\usepackage{graphicx}
\usepackage{epstopdf}
\usepackage{subfigure}
\newcommand{\goodgap}{%
	\hspace{\subfigtopskip}%
	\hspace{\subfigbottomskip}
}
\begin{document}	
	\author{Cristina Blaga\\Faculty of Mathematics and Computer Science \\Babe\c{s}-Bolyai University, Cluj-Napoca, Romania}
	\title{Timelike geodesics around a charged spherically symmetric dilaton black hole}
	\date{}
	\maketitle
	\begin{abstract}
In this paper we study the timelike geodesics around a spherically symmetric charged dilaton black hole. The trajectories around the black hole are classified using the effective potential of a free test particle. This qualitative approach enables us to determine the type of the orbit described by the test particle without solving the equations of motion, if the parameters of the black hole and the particle are known. The connections between these parameters and the type of orbit described by the particle are obtained. To visualize the orbits we solve numerically the equation of motions for different values of the parameters envolved in our analysis. The effective potential of a free test particle looks different for a non-extremal and an extremal black hole, therefore we have examined separately these two types of black holes.
\end{abstract}

\section{Introduction}

 In the classical general relativity the geometry of the spacetime near a charged black hole is described by Reissner-Nordstr\o m metric. In the low-energy string theory, the solution for the static charged black hole was obtained by Gibbons and Maeda (1988) and three years later, independently, by Garfinkle, Horowitz and Strominger (1991). Thus the static spherically symmetric charged dilaton black hole is known as Gibbons--Maeda--Garfinkle--Horowitz--Strominger (GMGHS) black hole. 
 
 In 1993 using a Harrison-like transformation (Harrison 1968) to the Schwarzschild solution, Horowitz (1993) derived a metric that fulfills Einstein-Maxwell field equations. For a massless dilaton the solution found by Horowitz corresponds to a GMGHS black hole. 
 
 The line element of the metric is:  
\begin{equation}\label{metr}
	ds^2=-\left(1-\frac{2M}{r}\right) dt^2 +\frac{1}{\left(1-\frac{2M}{r} \right)} d r^2
	+ r \left(r-\frac{Q^2}{M}\right) (d \theta^2 + \sin^2 \theta \, d \varphi^2)
\end{equation}
where $Q$ is related to the electrical charge of the black hole and $M$ to its mass. If $Q^2 < 2 M^2$ the singularity is inside the event horizon. If $Q^2 = 2 M^2$, the area of the events horizon shrinks to zero. This case is known as \emph{extremal limit}. 

The region inside the events horizon of a black hole cannot be seen by an external observer. All the information concerning the black hole are derived studying the behaviour of matter and light in its exterior. A free test particle moves on a timelike geodesics and a photon on a null geodesics. This is the reason why if we want to describe the motion in the vicinity of a black hole we have to solve the geodesic equations. 

In classical general relativity as well as in string theory, geodesics of black holes are studied extensively. The critical parameters for photon and timelike geodesics for Kerr black holes and Reissner-Nordstr\o m black holes have been found from conditions for multiple roots of corresponding polynomials by Zakharov (1994, 2014) and Zakharov et. al. (2005a,b, 2012).
The null geodesics of static charged black holes in heterotic string theory were analysed by Fernando (2012). The circular null and timelike geodesics for non-extremal and extremal GMGHS black hole were investigated by Pradhan (2012). The exact solutions of timelike geodesics of an extremal GMGHS black hole were derived in Blaga and Blaga (1998). The existence and stability of the circular orbits for the timelike geodesics of a GMGHS black hole were examined in Blaga (2013). A quantitative analysis of the timelike geodesics of a charged black hole in heterotic string theory was done by Olivares and Villanueva (2013), article in which the authors obtained the solution of the geodesic equations in terms of elliptic $\wp$-Weierstrass function.  

Our goal is to classify the orbits of free test particles in the vicinity of a spherically symmetric dilaton black hole using the effective potential. This approach enables us to identify the type of orbit described by a particle around a GMGHS black hole without solving the equations of motion. The type of the orbit described by a free test particles around a given GMGHS black hole depends on its angular momentum, energy and initial position. The qualitative results are accompanied by plots of possible orbits described by free test particles around different GMGHS, obtained by numerical integration of the equations of motion. The initial condition for a given type of black hole were chosen using the knowledge about the peculiarities of the effective potential.   

The paper is organized as follows: in the section 2 we derive the geodesic equations from the Lagrangian corresponding to the metric (\ref{metr}). In the section 3 we emphasize some properties of the effective potential of a free test particle for both a non-extremal and extremal GMGHS. In the section 4 we classify the possible trajectories around a GMGHS black hole with $b \in [0,2)$, respectively $b=2$ and plot the numerical solution of the geodesic equations for different values of the parameters involved in our analysis. The conclusion are given in the last section.

\section{Geodesic equations}

We can derive the geodesic equations directly computing the Christoffel coefficients for the given metric or using Hamilton-Jacobi theory (see for example Blaga and Blaga 1998) or writing the Euler-Lagrange equations (see Fernando 2012). We follow here the later approach described by Chandrasekhar (1983). The Lagrangian for the metric (\ref{metr}) is: 
\begin{equation}\label{lagr}
	2 \mathcal{L} = -\left(1-\frac{2M}{r}\right) \dot{t}^2 + \left(1-\frac{2M}{r} \right)^{-1} \dot{r}^2\\ 
	+ r \left(r-\frac{Q^2}{M}\right) \left(\dot{\theta}^2  + \sin^2 \theta \,\dot{\varphi}^2 \right) 
\end{equation}
where the dot means the differentiation with respect to $\tau$ - affine parameter along the geodesic. This parameter is chosen so that $2 \mathcal{L}=-1$ on a timelike geodesics, $2 \mathcal{L}=0$ on a null geodesics and $2 \mathcal{L}=1$ on a space-like geodesics.  
 
The coordinates $t$ and $\varphi$ are cyclic. Therefore the motion has two first integrals derived from the Euler-Lagrange for time and latitude  as follows. From the Euler-Lagrange equation for $t$ we get 
\begin{equation}\label{ien}
\left(1-\frac{2M}{r}\right) \dot{t} = \,E 
\end{equation}
known as the energy integral. The real constant $E$ is the total energy of the particle. 
The second integral of motion is obtained from the Euler-Lagrange for $\varphi$
\begin{equation}\label{imc}
2 r \left(r-\frac{Q^2}{M}\right) \sin^2\theta \, \dot{\varphi} = \mbox{constant}
\end{equation}
known as the integral of angular momentum. 

From the Euler-Lagrange equation for $\theta$ we get
\begin{equation}\label{ecteta}
\frac{d}{d \tau}\left[ r \left( r - \frac{Q^2}{M}  \right) \dot{\theta} \right] = r \left( r - \frac{Q^2}{M}  \right) \sin \theta \cos \theta \cdot \dot{\varphi}^2\,.
\end{equation}
Considering $\theta = \pi/2$ when $\dot{\theta}=0$ then $\ddot{\theta} = 0$ and $\theta=\pi/2$ on geodesic. This means that the motion is confined in a plane as in Schwarzschild spacetime or in Newtonian gravitational field.  

If $\theta=\pi/2$ during the motion the angular momentum integral (\ref{imc}) becomes 
\begin{equation}\label{ecL}
r \left(r-\frac{Q^2}{M}\right) \dot{\varphi} = L
\end{equation} 
where the real constant $L$ is the angular momentum about an axis normal at the plane in which the motion took place. 

The Euler-Lagrange equation corresponding to the radial coordinate is complicated. Therefore, we replace it with a relation derived from the constancy of the Lagrangian. If we substitute in Eq. (\ref{lagr}) $\dot{t}$ from Eq. (\ref{ien}), $\dot{\varphi}$ from Eq. (\ref{ecL}) and $\dot{\theta}$ from Eq. (\ref{ecteta}), after some algebra we get 
\begin{equation}\label{et}
\left( \frac{dr}{d \tau} \right)^2 + \left( 1 - \frac{2 M}{r} \right) \left(\frac{L^2}{r \left( r - \frac{Q^2}{M} \right)}  - \epsilon \right) = E^2
\end{equation}   
where $\epsilon=-1$ for timelike geodesics, $\epsilon=0$ for null geodesics and $\epsilon=+1$ for space-like geodesics. 

\subsection{Radial timelike geodesics}

The angular momentum of a free test particle moving on a radial timelike geodesics is zero. Therefore the Eq. (\ref{et}) becomes
\begin{equation}\label{gr}
\left( \frac{dr}{d \tau} \right)^2 + \left( 1 - \frac{2 M}{r} \right) = E^2\,.
\end{equation} 
If we want to study the motion of a test particle which describe a radial timelike geodesic we have to consider the Eq. (\ref{gr}) for radial coordinate $r$ and the Euler-Lagrange equation for time coordinate $t$ 
\begin{equation}\label{ient}
\frac{d t}{d \tau} =\frac{E}{\left(1-\frac{2M}{r}\right)}\,, 
\end{equation}
where $\tau$ is the proper time. These two equations are identical with the equations of radial timelike geodesics in Schwarzschild spacetime (Chandrasekhar 1983). Therefore, the motion of free test particles on radial geodesics in the vicinity of these two kind of black holes will have the same properties. It is well known that an observer stationed at a large distance from a Schwarzschild black hole finds that a free falling test particle approaches to the events horizon but can never cross it, because the test particle needs an infinite time $t$ to reach the horizon, even though it crosses the horizon in a finite proper time $\tau$. This property is valid for a free falling particle on a radial timelike geodesic of GMGHS black holes, too.  

\section{Effective potential for timelike geodesics}

If we compare Eq. (\ref{et}) with $\dot{r}^2+V_{\mbox{\scriptsize{eff}}}=E^2$ we get the effective potential    
\begin{equation}\label{Veff}
	V_{\mbox{\scriptsize{eff}}} = \left( 1 - \frac{2 M}{r} \right) \left(\frac{L^2}{r \left( r - \frac{Q^2}{M} \right)}  - \epsilon \right) 
\end{equation}
which depends on radial coordinate $r$, the parameters related to the electrical charge and mass of the black hole, the type of the geodesics and the angular momentum of the particle. 

For timelike geodesics $\epsilon=-1$ and (\ref{Veff}) becomes
\begin{equation}\label{Veft}
	V_{\mbox{\scriptsize{eff}}} = \left( 1 - \frac{2 M}{r} \right) \left(\frac{L^2}{r \left( r - \frac{Q^2}{M} \right)}  +1 \right). 
\end{equation}
In order to lower the number of the parameters from the effective potential we denote $u=r/M$, $a=(L/M)^2$, $b=(Q/M)^2$ and the relation (\ref{Veft}) becomes 
\begin{equation}\label{Vefu}
	V_{\mbox{\scriptsize{eff}}} = \left( 1 - \frac{2}{u} \right) \left(\frac{a}{u \left( u - b \right)} + 1 \right) \,.
\end{equation}
The quantity $\mathit{q}=Q/M$ is the specific electrical charge and we observe that $b=\mathit{q}^2$. In this paper we have assumed that $Q^2 \leq M^2$ hence $b \in \left[0,2 \right]$, where $b=0$ is for a Schwarzschild black hole and $b=2$ for an extremal GMGHS black hole. The parameter $a$ equals $(L/M)^2$, therefore in our study it will be positive or zero. In the new variable $u$, the events horizon corresponds to $u=2$ and the region outside the horizon will be given by $u > 2$.

\subsection{Effective potential for $b \in [0,2)$}

\begin{figure}
	\centering
	\subfigure[]{\label{fig:fig1-a}{\includegraphics[width=.38\textwidth]{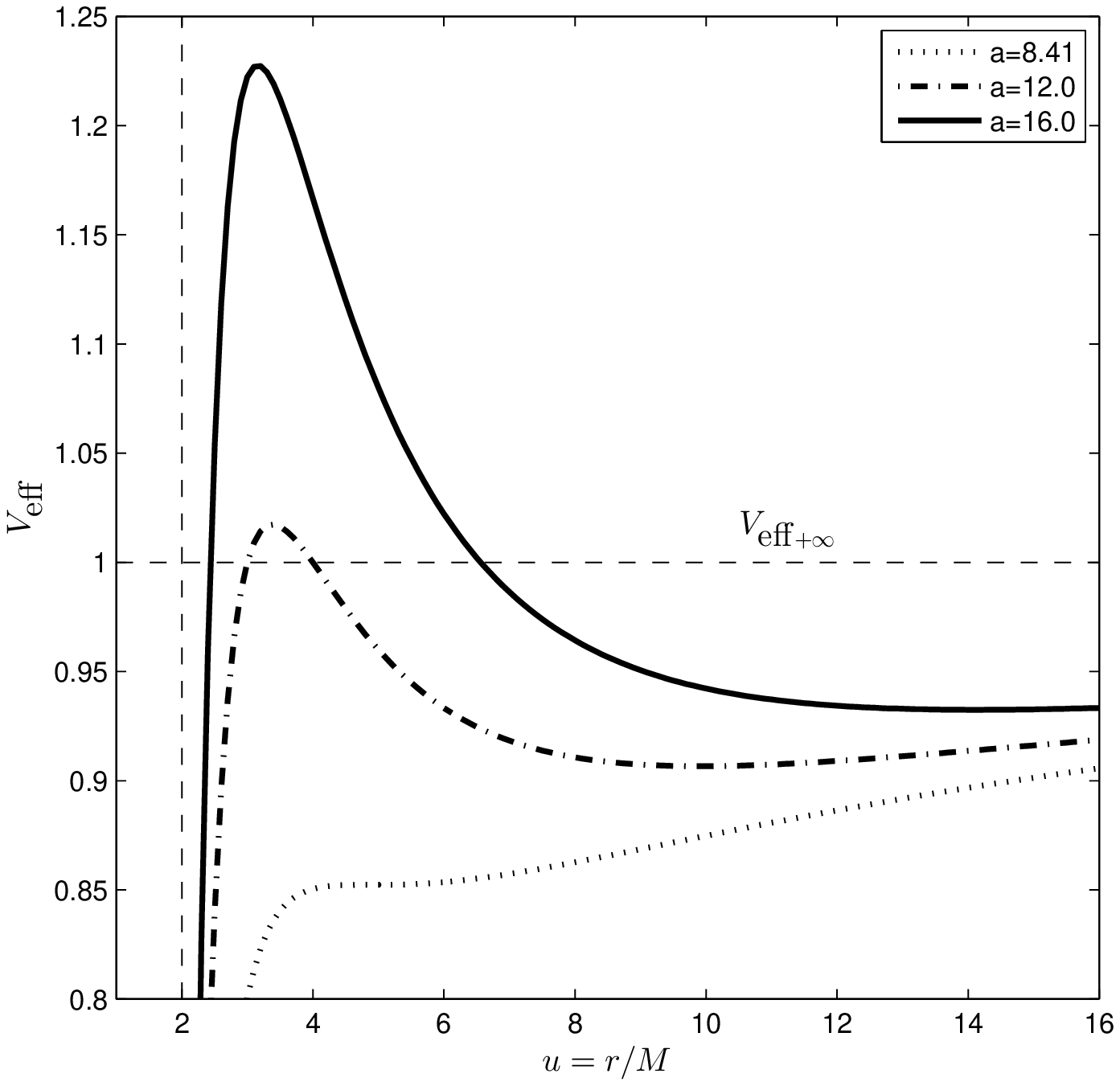}}}\goodgap
	\subfigure[]{\label{fig:fig1-b}{\includegraphics[width=.38\textwidth]{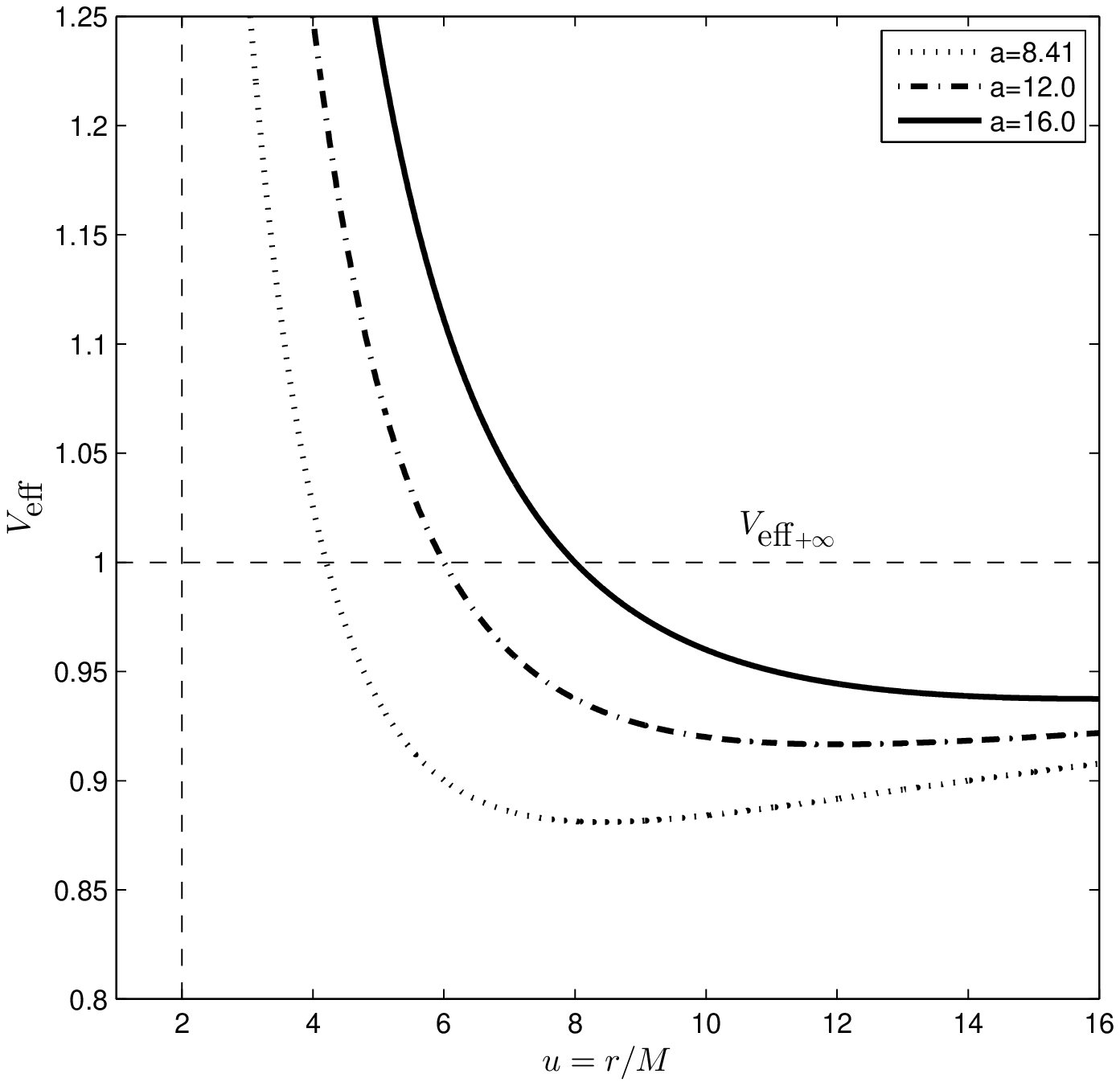}}}
	\caption{\subref{fig:fig1-a} Effective potential for $b=1$ and $a \in  \{8.41,12,16\}$. The dashed lines are for the vertical $u=2$ - events horizon and horizontal $V_{\mbox{\scriptsize{eff}}}=1$ - the limit of the potential as $u \rightarrow +\infty$.
		\subref{fig:fig1-b} Effective potential for $b=2$ and $a \in  \{8.41,12,16\}$. The dashed lines are for the vertical $u=2$ - events horizon and horizontal $V_{\mbox{\scriptsize{eff}}}=1$ - the limit of the potential as $u \rightarrow +\infty$.}
	\label{fig:fig1}
\end{figure}

The effective potential (\ref{Vefu}) was studied in Ref. \cite{b13}. The qualitative analysis of $V_{\mbox{\scriptsize{eff}}}$ revealed that for $b \in [0,2)$, the function is positive for $u \in [2, +\infty)$, it is zero on the events horizon $u=2$ and approaches $1$ as $u$ approaches $+\infty$ (see Fig. \ref{fig:fig1-a}). If $a$ is greater then a value depending on $b$ (see \cite{b13}), the potential has two local extreme points outside the events horizon, the local maximum point of $V_{\mbox{\scriptsize{eff}}}$ being closest to the horizon. For a given $b$, if there is a local maximum point outside the horizon, the maximum value of the potential increases with $a$. 

In Fig. \ref{fig:fig1-a} we have plotted the effective potential against $u=r/M$ for $b=1$ and three values of $a$. The first value $a=8.41054$ is the lowest value of $a$ for which $V_{\mbox{\scriptsize{eff}}}$ admits extreme points outside the horizon for $b=1$ (see \cite{b13}). The second value $a=12$ is the lowest value of $a$ for which the effective potential admits extreme points outside the horizon in the case of a Schwarzschild black hole $b=0$ (see \cite{cha}). 

If we choose another value for $b \in [0,2)$ different from $1$, the graph of the effective potential is similar, but for a given value of $a$, if the potential admits two extreme points outside the events horizon, the maximum value of the potential increases with $b$ (see \cite{b13}). 

\subsection{Effective potential for $b=2$}

For a test particle moving around an extremal GMGHS black hole $b=2$ the effective potential is
\begin{equation}\label{Vefb2}
	V_{\mbox{\scriptsize{eff}}} (u) = \frac{u^2-2 u+a}{u^2}\,,
\end{equation}  
function which is not defined in $u=0$, approaches infinity as $u$ approaches zero and $1$ as $u$ approaches infinity (see Fig. \ref{fig:fig1-b}). In this case the effective potential admits only a local minimum point in $u=a$. This point is outside the events horizon if and only if $a \geq 2$. The minimum value of the effective potential increases with $a$, it approaches $1$ from below as $a$ approaches $+\infty$. We have plotted the effective potential (\ref{Vefb2}) for various values of $a$ in Fig. \ref{fig:fig1-b}. 

\section{Types of timelike geodesics}

If we replace $\varepsilon=-1$ in Eq. (\ref{et}) we get the equation of motion of a free test particle around a spherically symmetric charged black hole. Therefore, the geometry of the orbit described by the test particle depends on its energy. 

The motion is possible only if $E^2 - V_{\mbox{\scriptsize{eff}}}(u) \geq 0$. The points in which $E^2 = V_{\mbox{\scriptsize{eff}}}(u)$ are known as turning points, because there ${d r}/{d t}=0$ and the direction of the motion could be changed. 

\subsection{Orbit around a dilaton black hole with $b \in [0,2)$}

\begin{figure}
	\centering
	\includegraphics[width=.38\textwidth]{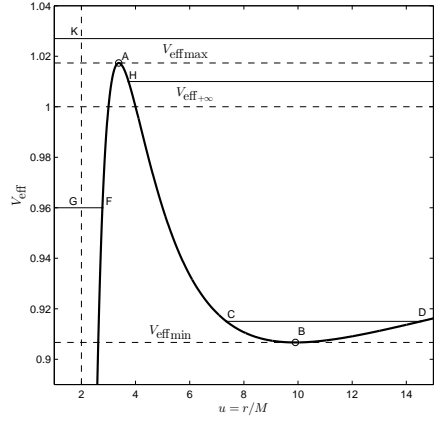}%
	\caption{\label{fig3} Effective potential for $b=1$ and $a=12$. Dashed lines are for the extreme values of the effective potential, the limit of the potential as $u$ approaches infinity ${V_{\mbox{\scriptsize{eff}}}}_{+\infty}$ and the vertical $u=2$. The solid lines are for different energy levels.}
\end{figure}

In Fig. \ref{fig3} we have plotted the effective potential for a dilaton black hole with $b=1$ as an exponent for GMGHS black holes with $b \in [0,2)$. The parameter $a$ equals $12$. The points $A$ and $B$ represent the maximum, respectively the minimum point of the effective potential. The dashed horizontal lines are for the maximum value of the potential ${V_{\mbox{\scriptsize{eff}}}}_{\mbox{\scriptsize{max}}}$, the minimum value ${V_{\mbox{\scriptsize{eff}}}}_{\mbox{\scriptsize{min}}}$ and the limit of the potential as $u$ approaches infinity ${V_{\mbox{\scriptsize{eff}}}}_{+\infty}$. The horizontal solid lines represent different energy levels. We have denoted with $C$, $D$, $F$ and $H$ the intersection points of an energy level with the graph of the effective potential. The points in which a test particle crosses the line $u=2$ are denoted with $G$ and $K$.          

If we consider a free test particle moving around a dilaton black hole with $b \in [0,2)$ and $a$ greater then the lower value for which the effective potential has extrema outside the events horizon, the graph of the effective potential is similar with the graph from Fig. \ref{fig3}. This is the reason why we have computed the orbits around a GMGHS black hole with $b=1$ to visualize the geometry of possible trajectories around GMGHS black holes with an arbitrary $b \in [0,2)$. In the following polar plots we have represented the region inside events horizon as a circle filled with black. The orbits were obtained by numerical integration of Eq. (\ref{et}) and Eq. (\ref{ecL}) using \textsc{Matlab}.        

If the motion of a free test particle around a dilaton black hole is possible, then depending on the total energy of the test particle and its initial position $u_0$, we can find the following qualitative motions:

\textbf{Case 1}: if $E^2 > {V_{\mbox{\scriptsize{eff}}}}_{\mbox{\scriptsize{max}}}$ then the test particle approaches to the black hole, crosses the events horizon and end into the singularity. The particle is gravitationally captured. 

\begin{figure}
	\centering
	\subfigure[]{\label{fig:fig3-a}{\includegraphics[width=.30\textwidth]{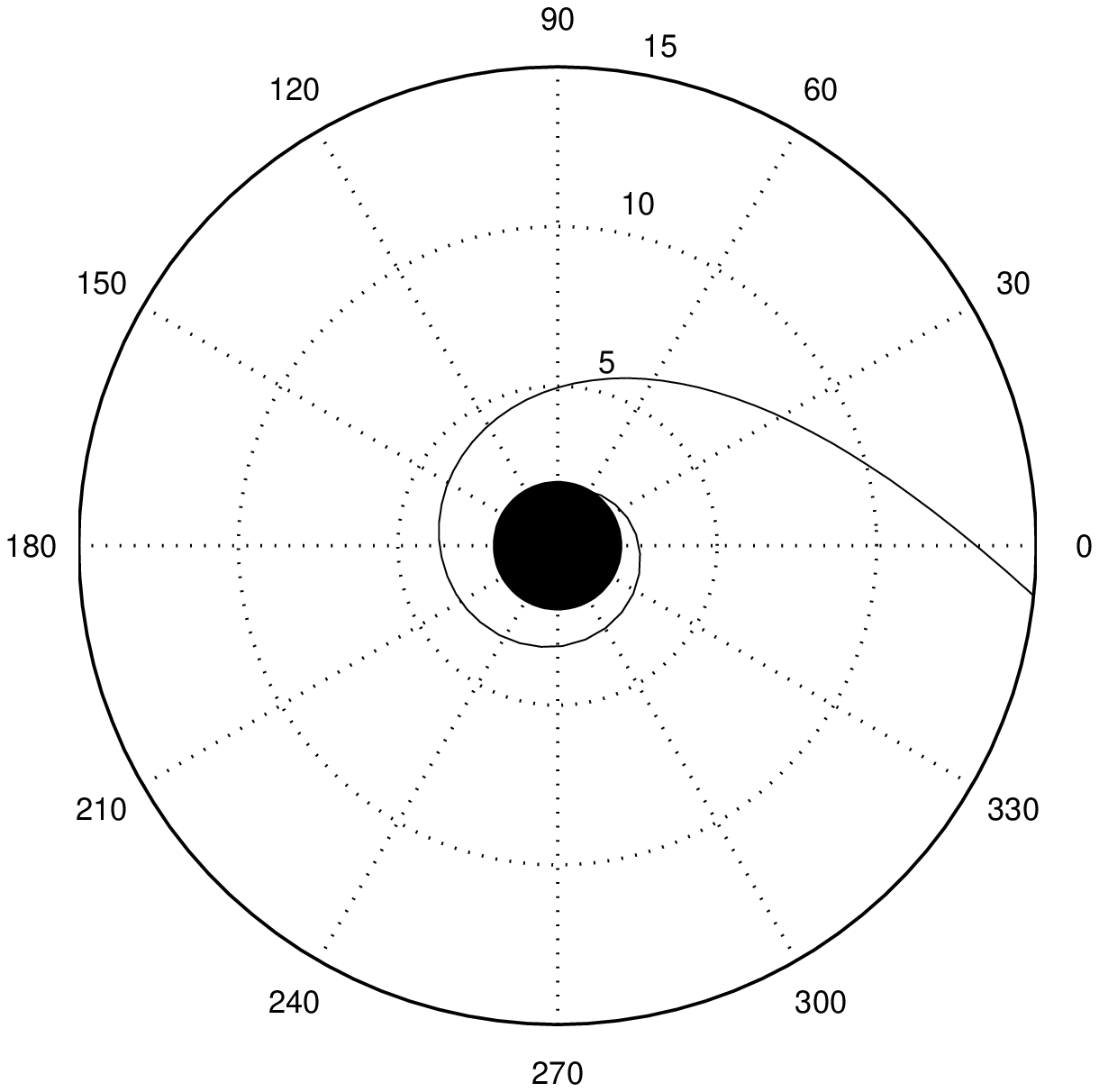}}}\goodgap
	\subfigure[]{\label{fig:fig3-b}{\includegraphics[width=.30\textwidth]{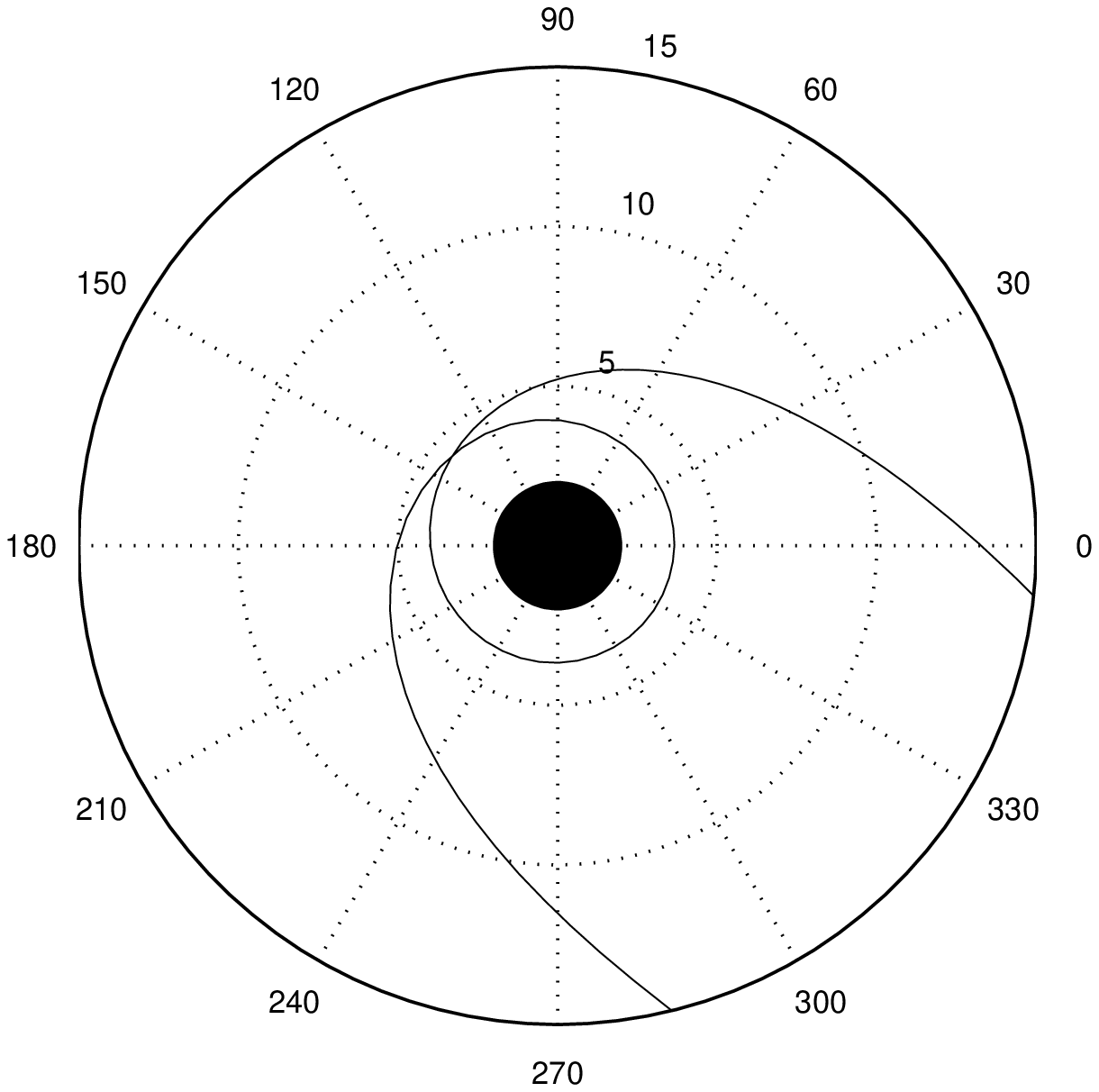}}}\\
	\subfigure[]{\label{fig:fig3-c}{\includegraphics[width=.30\textwidth]{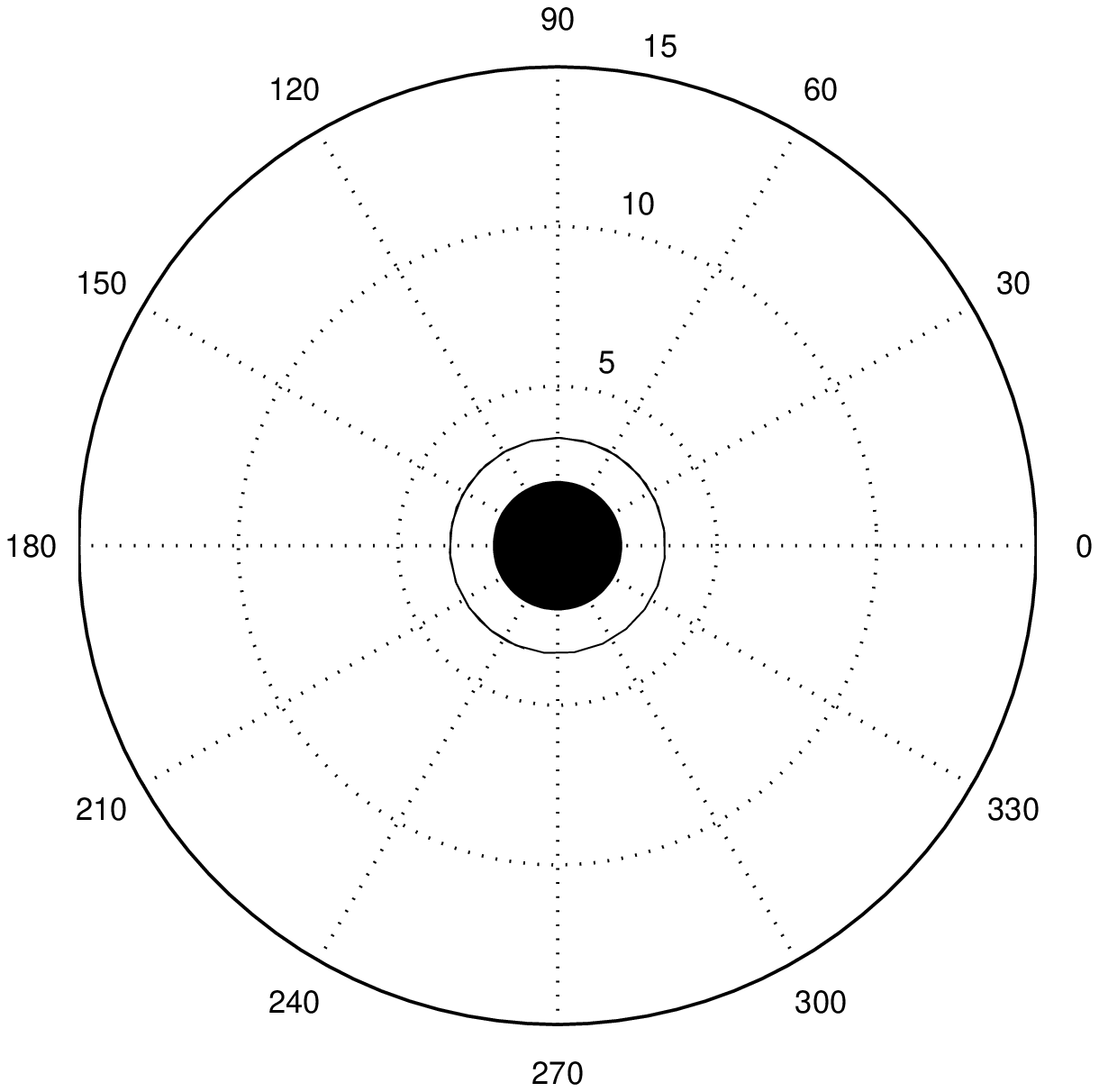}}}\goodgap
	\subfigure[]{\label{fig:fig3-d}{\includegraphics[width=.30\textwidth]{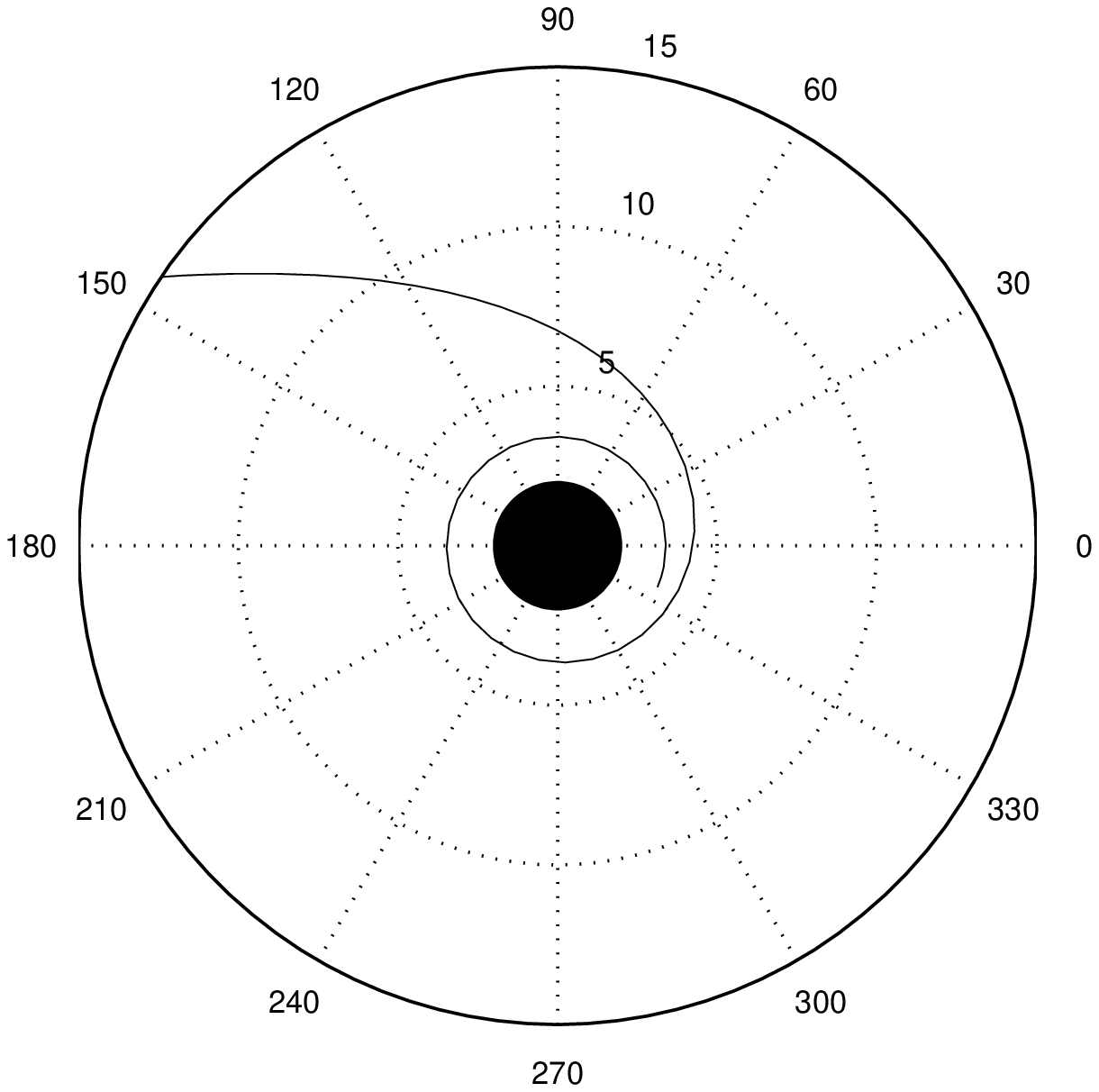}}}\\
	\subfigure[]{\label{fig:fig3-e}{\includegraphics[width=.30\textwidth]{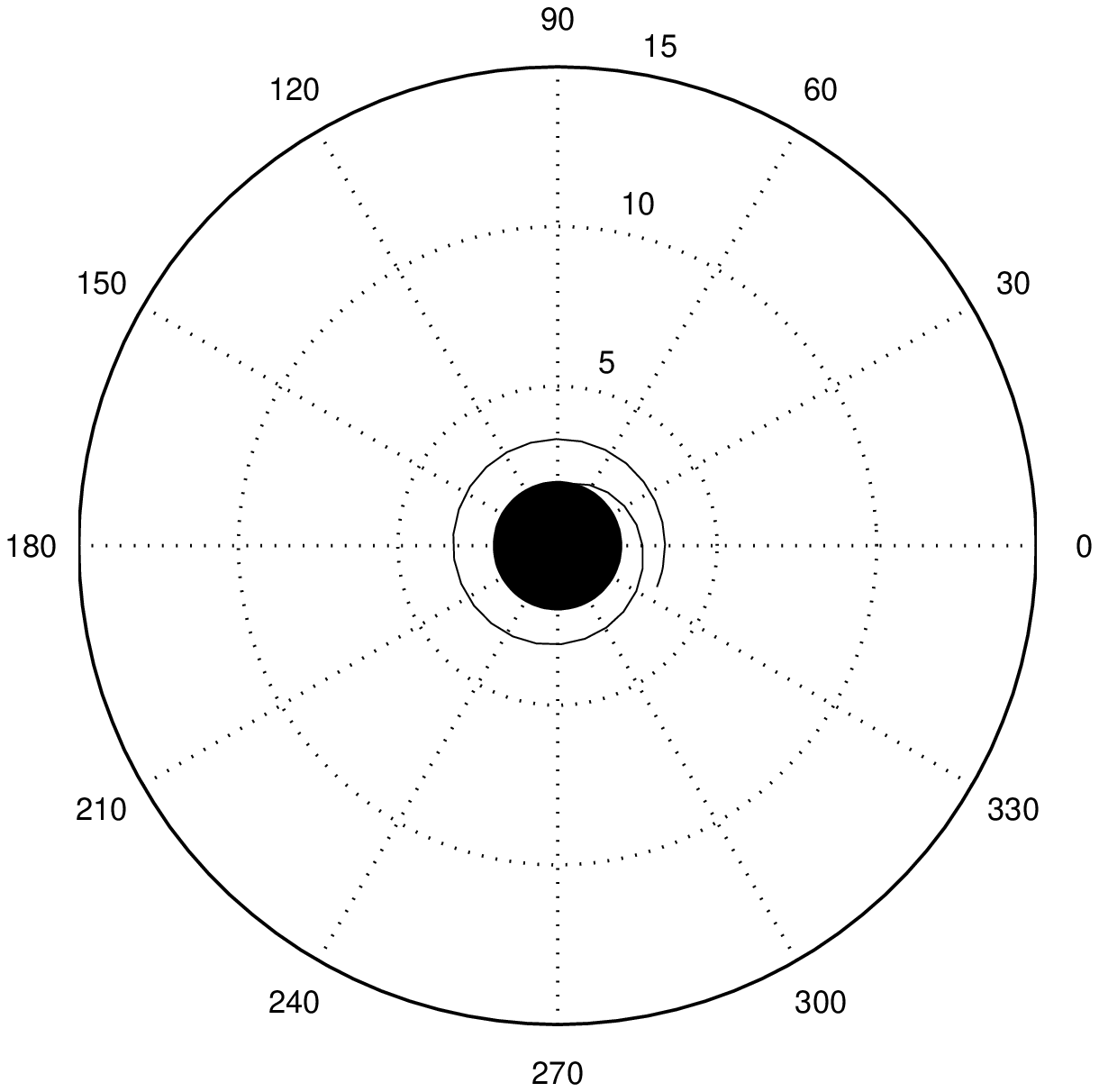}}}\goodgap
	\subfigure[]{\label{fig:fig3-f}{\includegraphics[width=.30\textwidth]{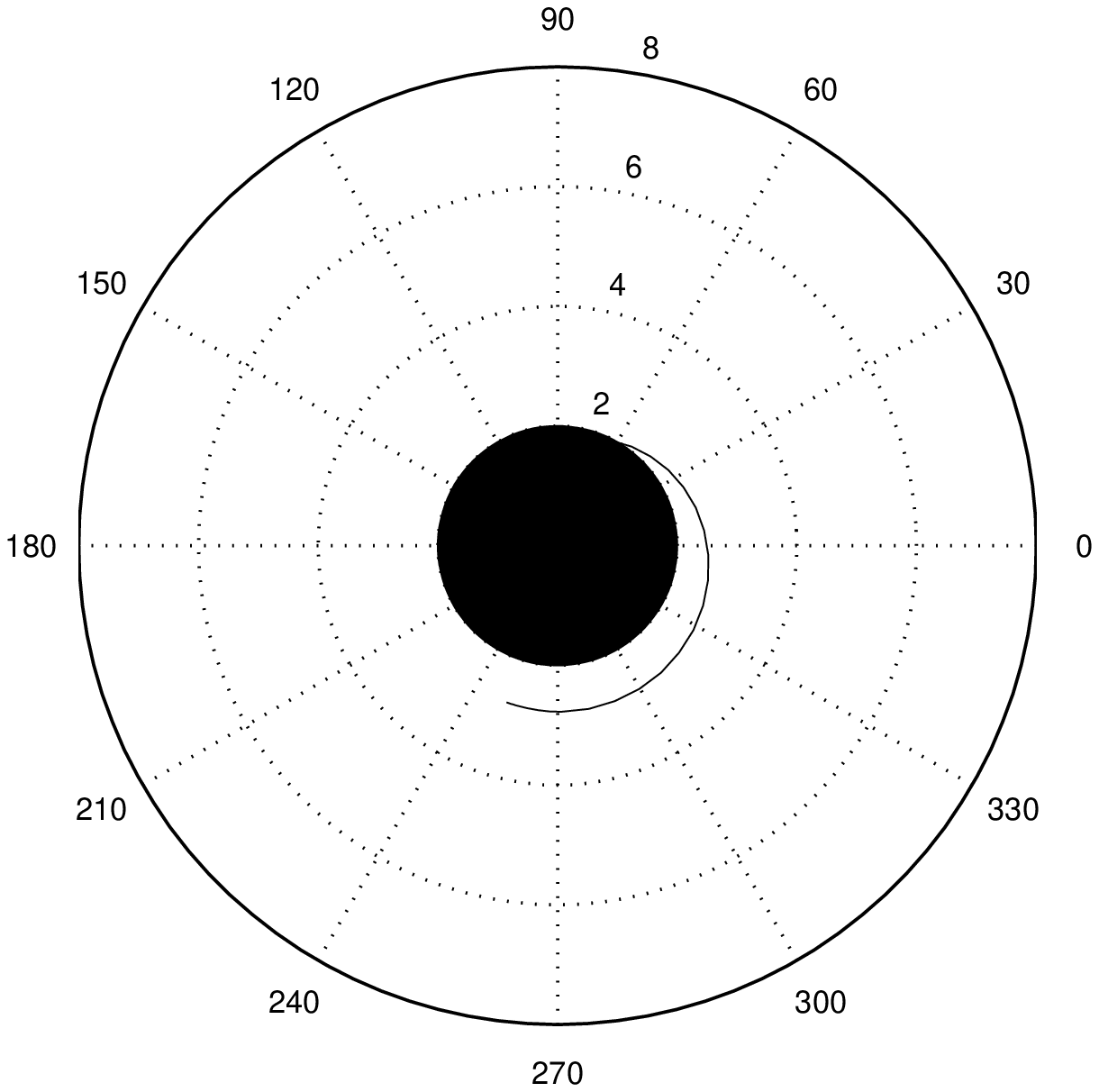}}}\\
	\subfigure[]{\label{fig:fig3-g}{\includegraphics[width=.30\textwidth]{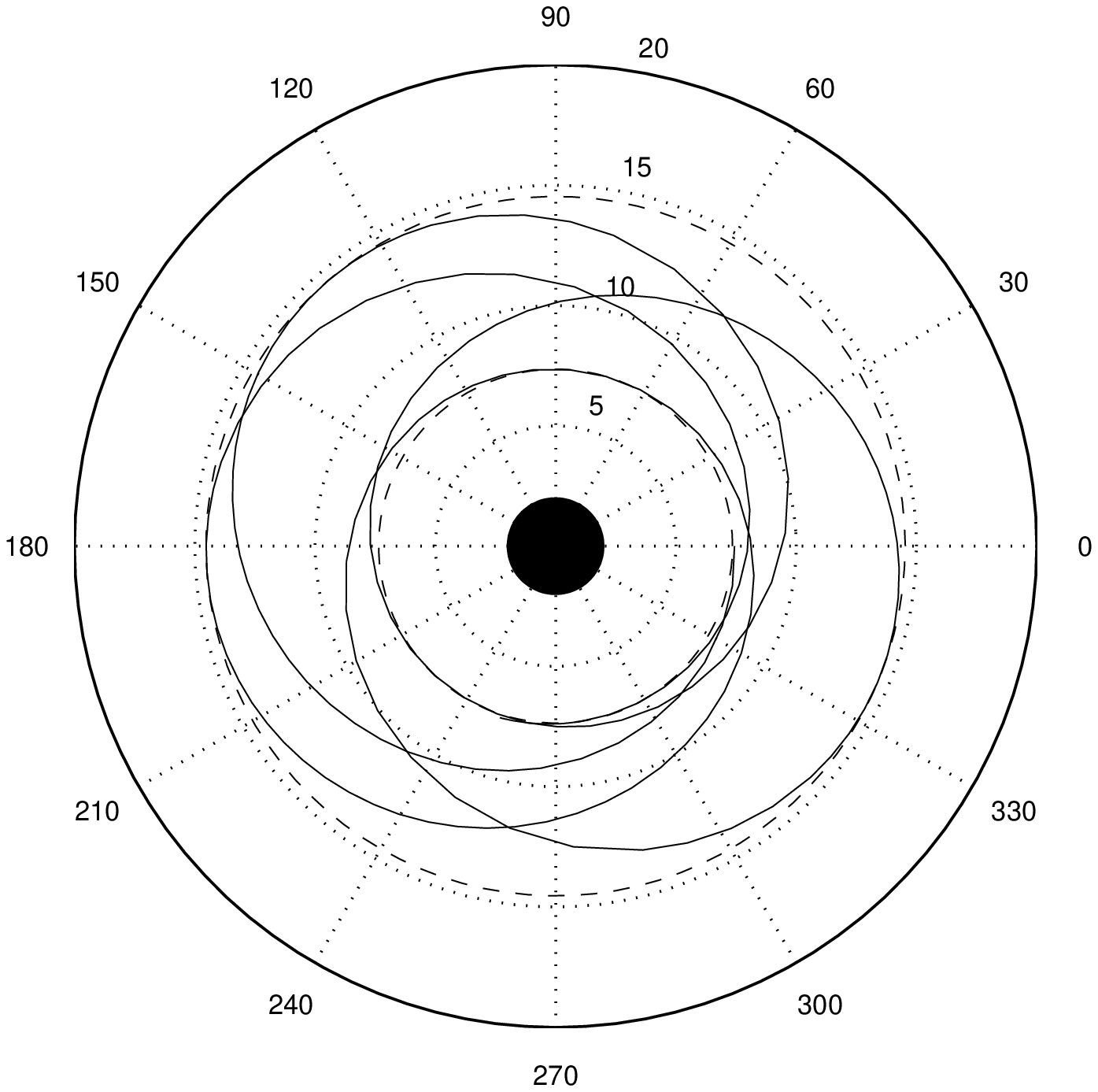}}}\goodgap
	\subfigure[]{\label{fig:fig3-h}{\includegraphics[width=.30\textwidth]{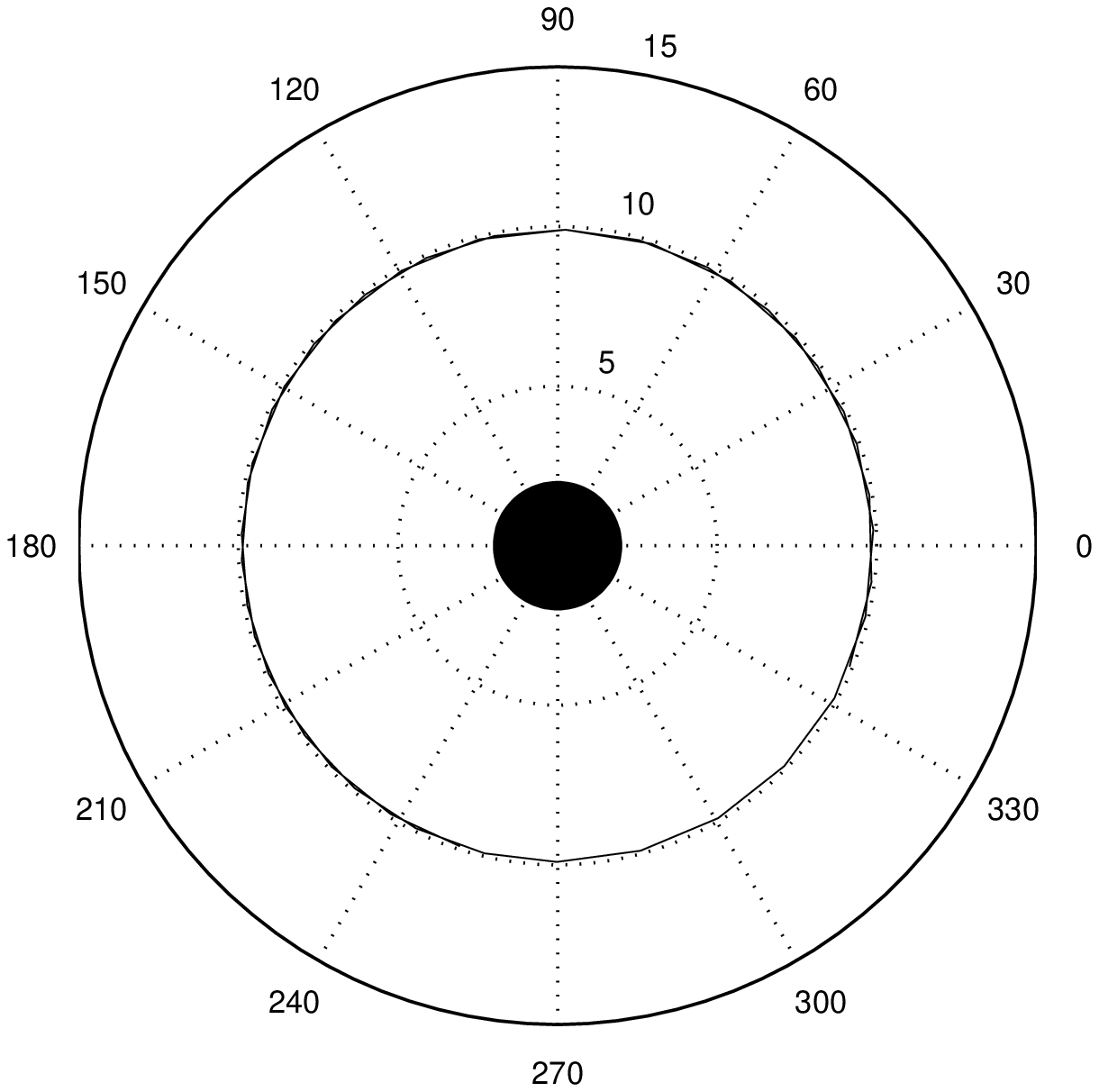}}}
	\caption[Orbits around a dilaton black hole with $b=1$ and $a=12$.]{Orbits around a dilaton black hole with $b=1$ and $a=12$. \subref{fig:fig3-a} Gravitational capture $E^2=1.03$, $u_0=15.0$. \subref{fig:fig3-b} Hyperbolic motion $E^2=1.01$, $u_0=15.0$. \subref{fig:fig3-c} Unstable circular motion $E^2=1.073$, $u_0=3.381$. \subref{fig:fig3-d} Outer marginal orbit $E^2=1.073$, $u_0=3.39$. \subref{fig:fig3-e} Inner marginal orbit $E^2=1.073$, $u_0=3.37$. \subref{fig:fig3-f} Motion trapped near horizon $E^2=0.96$, $u_0=2.75$. \subref{fig:fig3-g} Bounded orbit $E^2=0.915$, $u_0=7.5$, $u(C)=7.355$, $u(D)=14.53$. \subref{fig:fig3-h} Stable circular orbit $E^2=0.907$, $u_0=9.9$.}
	\label{fig:fig3}
\end{figure}

In Fig. \ref{fig:fig3-a} we have represented a gravitational capture. The free test particle is moving with the total energy corresponding to the highest energy from Fig. \ref{fig3}, crosses the events horizon in $K$ and end into the singularity. 

\textbf{Case 2}: if $E^2 \in (1,{V_{\mbox{\scriptsize{eff}}}}_{\mbox{\scriptsize{max}}})$, $u_0 > u(A)$ and the motion is possible the test particle approaches the black hole until its squared energy equals the potential and then goes back to infinity (see Fig. \ref{fig:fig3-b}). We say that the particle is scattered hyperbolically or it has an hyperbolic motion. 

In Fig. \ref{fig:fig3-b} we have represented the orbit of a test particle with an energy equal to the energy from the second energy level from Fig. \ref{fig3}. In $H$ the test particle change its direction of motion and turns back to large distances from the black hole.    

\textbf{Case 3}: if $E^2 = {V_{\mbox{\scriptsize{eff}}}}_{\mbox{\scriptsize{max}}}$ and $u_0=u(A)$ then the test particle describes a circular orbit around the black hole (see Fig. \ref{fig:fig3-c}). Since $A$ is a local maximum point for effective potential the circular orbit is unstable.   

\textbf{Case 4}: if $E^2 = {V_{\mbox{\scriptsize{eff}}}}_{\mbox{\scriptsize{max}}}$, $u_0 > u(A)$ and the motion is possible, the particle escapes to infinity. We say that it is moving on an outer marginal orbit (see Fig. \ref{fig:fig3-d}). If $E^2 = {V_{\mbox{\scriptsize{eff}}}}_{\mbox{\scriptsize{max}}}$ and $u_0 < u(A)$, the particle ends into the singularity and we say that it describes an inner marginal orbit (see Fig. \ref{fig:fig3-e}). 

\textbf{Case 5}: if $E^2 < {V_{\mbox{\scriptsize{eff}}}}_{\mbox{\scriptsize{max}}}$, $u_0 < u(A)$ and the motion is possible then the test particle moves toward the singularity crossing the horizon in $F$ (see Fig. \ref{fig:fig3-f}). The motion is trapped near horizon.  

\textbf{Case 6}: if $E^2 \in ({V_{\mbox{\scriptsize{eff}}}}_{\mbox{\scriptsize{min}}},\min( 1, {V_{\mbox{\scriptsize{eff}}}}_{\mbox{\scriptsize{max}}}))$, $u_0>u(A)$ and the motion is possible then the test particle moves toward and backward from the black hole. $C$ and $D$ are the points in which the test particle change its direction of motion. The orbit is bounded and the motion is restricted in the annular region placed between the circles with radius $u(C)$ and $u(D)$ respectively.  

In Fig. \ref{fig:fig3-g} we have plotted the orbit of a free test particle with energy corresponding to the lowest energy level from Fig. \ref{fig3}. The dashed lines are for the circles with radius $u=u(C)$ - the lower limit of the distance between the test particle and the singularity ($u=0$), respectively $u=u(D)$ - the upper limit of the same distance.

\textbf{Case 7}: if $E^2 = {V_{\mbox{\scriptsize{eff}}}}_{\mbox{\scriptsize{min}}}$ and $u_0=u(B)$ the test particle describes a stable circular orbit, because $B$ locates a minimum point of the effective potential (see Fig. \ref{fig:fig3-h}).

The unstable and stable circular orbits are in fact special bounded orbits with $u=u(A)$ and $u=u(B)$ respectively. 

We observe that the type of trajectories enumerated in this section are qualitative similar with the trajectories found in a Schwarzschild spacetime (see Ref~\cite{fz}). This fact was expected because a Schwarzschild black hole is a GMGHS black hole with $b=0$. The difference between the orbits around a Schwarzschild black hole and a GMGHS dilaton black hole with $b \in (0,2)$ appears in the dynamical parameters of the motion. For example, for a given $a$ the test particle needs an higher energy to be gravitationally captured, because the maximum value of the effective potential increases with $b$.

\subsection{Trajectories around an extremal GMGHS black hole}

\begin{figure}
	\centering
	\includegraphics[width=.38\textwidth]{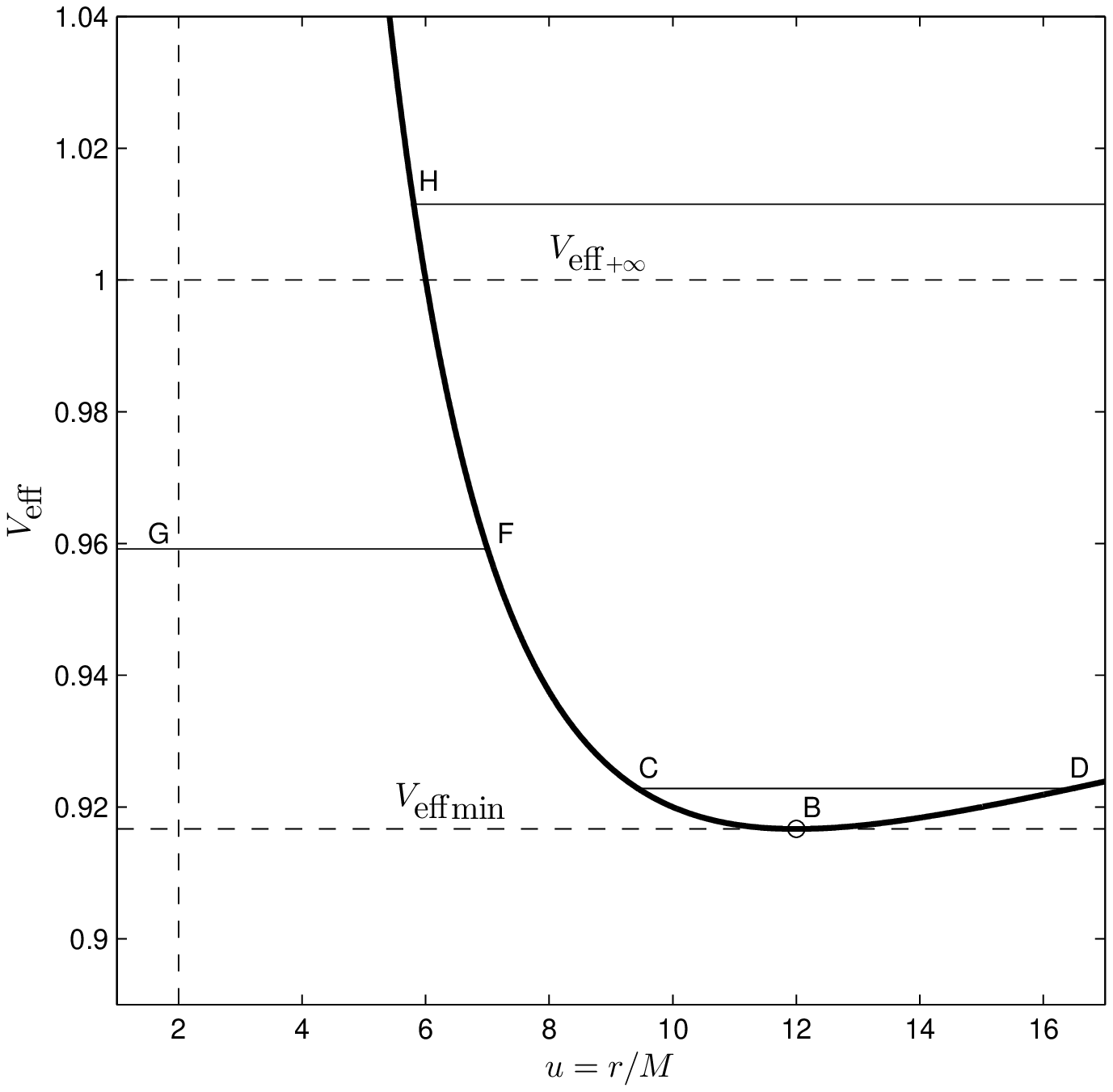}%
	\caption{\label{fig12} Effective potential for $b=2$. The dashed lines are for the minimum value of the effective potential, the limit of the potential as $u$ approaches infinity ${V_{\mbox{\scriptsize{eff}}}}_{+\infty}$ and the vertical $u=2$. The solid lines are for different energy levels.}%
\end{figure}

In Fig. \ref{fig12} we have plotted the effective potential for an extremal dilaton black hole for $a=12$. The horizontal dashed lines are for minimum value of the effective potential and its limit when $u \rightarrow +\infty$. The vertical $u=2$ plotted with dashed line corresponds to the events horizon. The energy levels are represented with solid lines. The point $B$ is the minimum point of the effective potential, $C$, $D$, $F$ and $H$ are the turning points and $G$ the point in which the test particles cross the horizon.   

\begin{figure}
	\centering
	\subfigure[]{\label{fig:fig5-a}{\includegraphics[width=.30\textwidth]{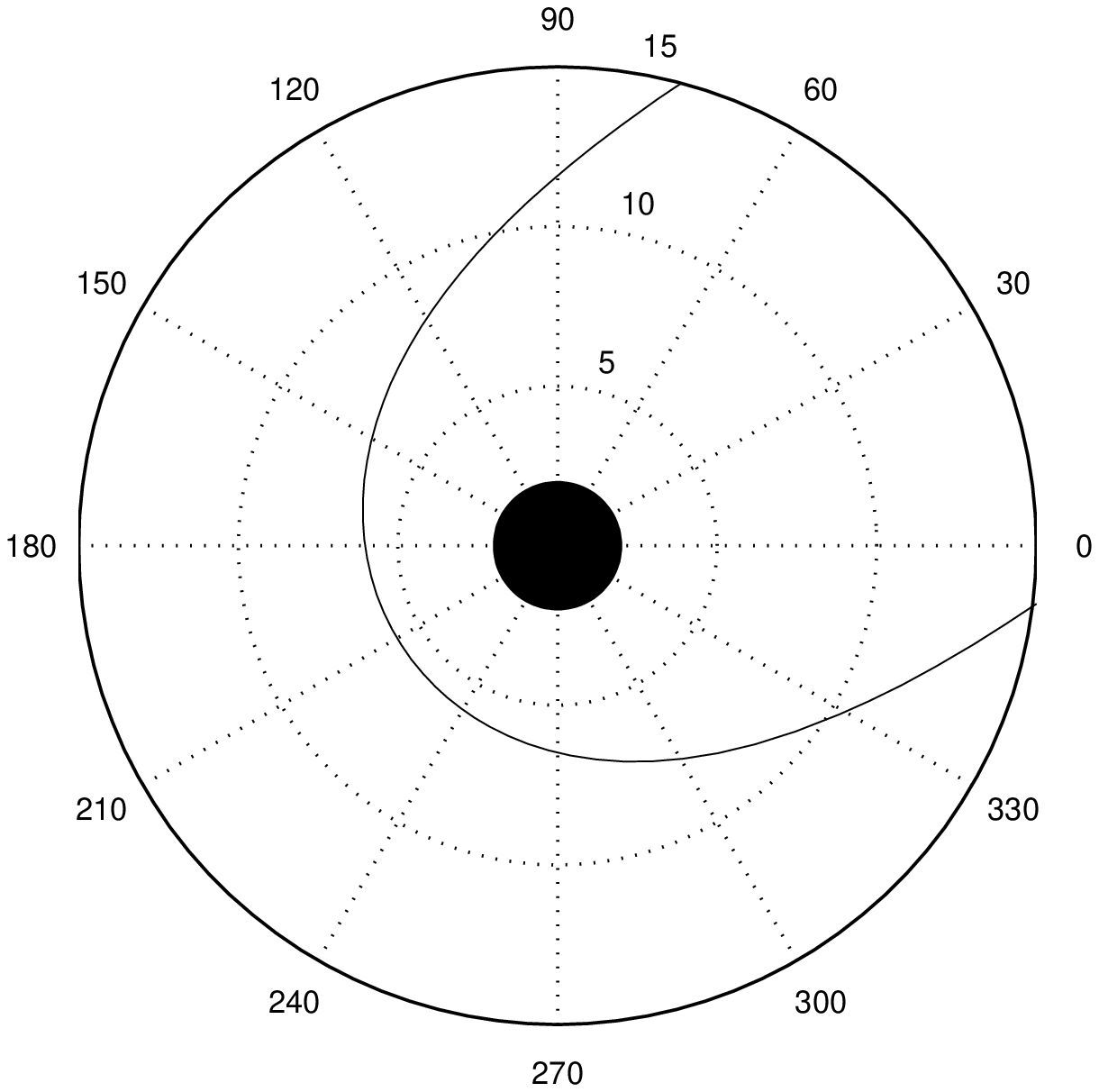}}}\goodgap
	\subfigure[]{\label{fig:fig5-b}{\includegraphics[width=.30\textwidth]{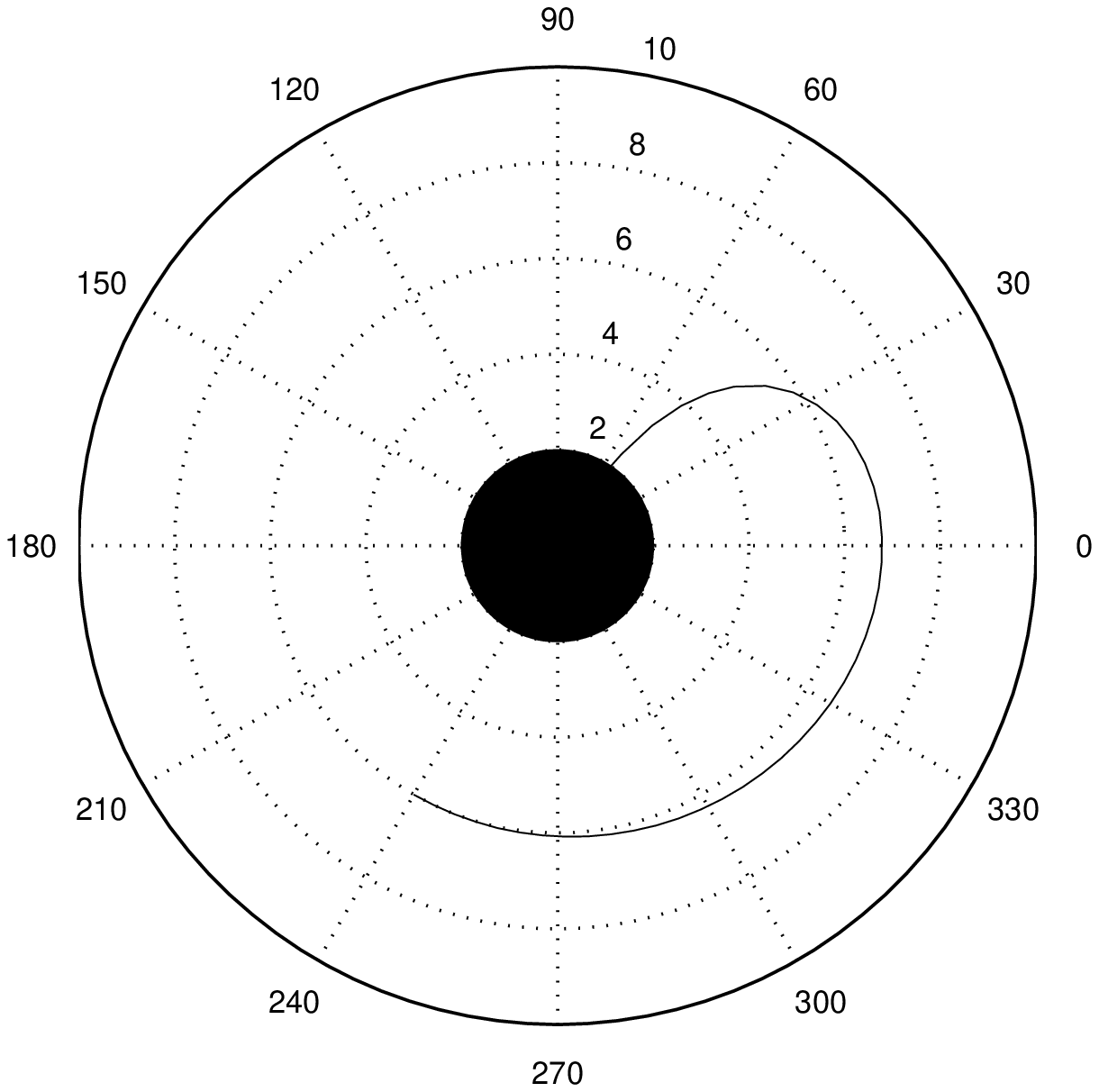}}}\\
	\subfigure[]{\label{fig:fig5-c}{\includegraphics[width=.30\textwidth]{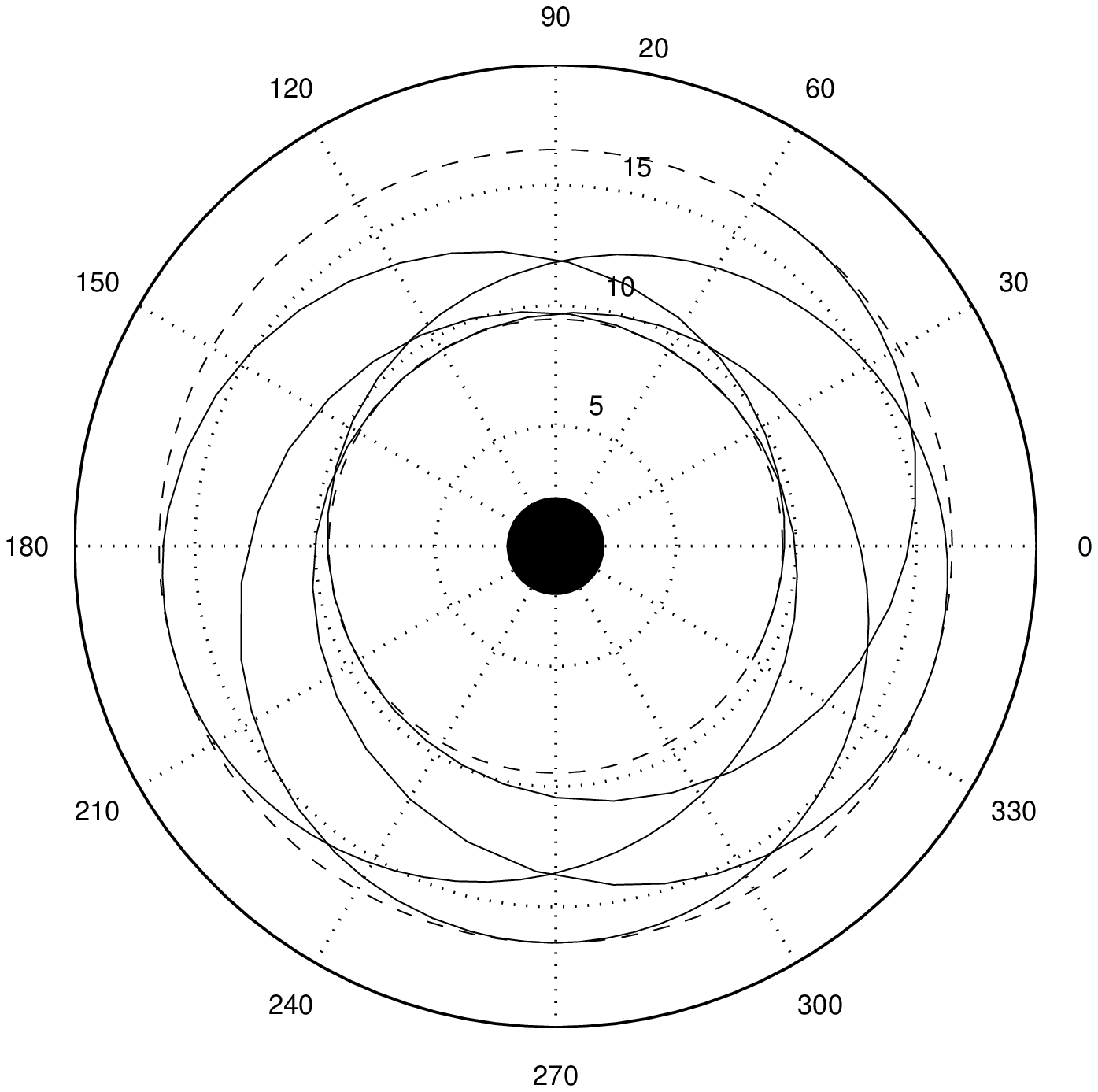}}}\goodgap
	\subfigure[]{\label{fig:fig5-d}{\includegraphics[width=.30\textwidth]{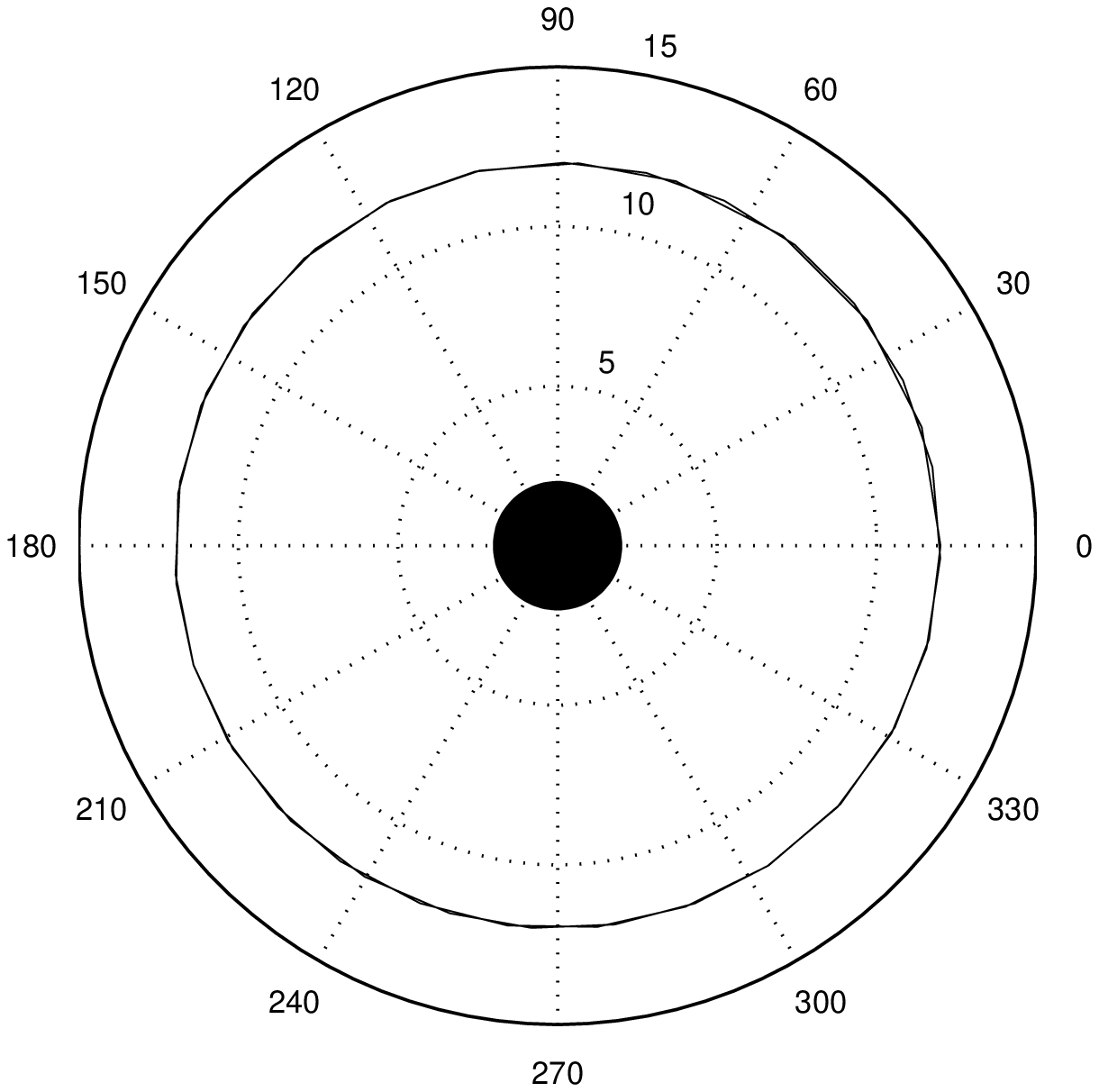}}}\\
	\caption[Orbits around a dilaton black hole with $b=2$ and $a=12$.]{Orbits around a dilaton black hole with $b=2$ and $a=12$. \subref{fig:fig5-a} Hyperbolic scattering $E^2=1.012$, $u_0=15.0$.\subref{fig:fig5-b} Near horizon trapped motion $E^2=0.96$, $u_0=6.0$. \subref{fig:fig5-c} Bounded orbit $E^2=0.922$, $u_0=9.45$. The dashed lines are for the circles $u(C)=9.43$ and $u(D)=16.49$. \subref{fig:fig5-d} Stable circular orbit $E^2=0.917$, $u_0=12.0$.}
	\label{fig:fig5}
\end{figure}

If we consider a free test particle moving around an extremal dilaton black hole $b = 2$ and $a \geq 2$, then the graph of the effective potential is similar with the plot from Fig. \ref{fig12}. Depending on the total energy of the particle and its initial position $u_0$ a free test particle describes one of the following trajectory: 

\textbf{Case 1}: if $E^2 \geq 1$ and $u_0 > u(H)$ where $u(H)>2$ is a solution of the equation $V_{\mbox{\scriptsize{eff}}}(u)=E^2$ the particle will move toward to the black hole, arrives in $H$ and goes back to infinity (see Fig. \ref{fig:fig5-a}). We say that the test particle has an hyperbolic motion, or is hyperbolically scattered. In Fig. \ref{fig:fig5-a} we have represented the hyperbolic motion of a test particle with the energy corresponding to the highest energy level from Fig \ref{fig12}.

\textbf{Case 2}: if $E^2 > {V_{\mbox{\scriptsize{eff}}}}_{\mbox{\scriptsize{min}}}$ and $u=r/M_0 < u(F)$ where $F$ is a solution of the equation $V_{\mbox{\scriptsize{eff}}}(u)=E^2$ the particle fall into the singularity, crossing the horizon in $G$. In this case the orbit is trapped near events horizon. In Fig. \ref{fig:fig5-b} the total energy of the test particle corresponds to the second energy level from Fig. \ref{fig12}. 

\textbf{Case 3}: if $E^2 \in ({V_{\mbox{\scriptsize{eff}}}}_{\mbox{\scriptsize{min}}},1)$ and $u_0>2$ the particle moves toward and backward from the black hole. The particle changes its direction of motion in the turning points $C$ and $D$ respectively (see Fig. \ref{fig:fig5-c}). In this case the motion is bounded. The radius of the limit circles are the solutions $u>2$ of the equation $V_{\mbox{\scriptsize{eff}}}(u)=E^2$. 

\textbf{Case 4}: if $E^2 = {V_{\mbox{\scriptsize{eff}}}}_{\mbox{\scriptsize{min}}}$ and $u_0=u(B)$ the particle describes a stable circular orbit, because $B$ is the minimum point of the effective potential (see Fig. \ref{fig:fig5-d}).

We observe that a free test particle is moving near an extremal dilaton black hole on trajectories qualitatively similar with those described in a Newtonian field. This similarity came from the fact that the effective potential (\ref{Vefb2}) is alike the effective potential from the motion in a Newtonian field (see Frolov and Zelnikov 2011). 

\section{Conclusions}

In this work we have considered the timelike geodesics of static spherically symmetric charged black hole known as GMGHS black holes. 
We have classified the solutions of the timelike geodesic equations using the effective potential of a free test particle moving in the vicinity of such a black hole. 

The effective potential of a test particle depends on radial coordinate $r$, the parameter related to the electrical charge $Q$, the mass $M$ of the black hole and the angular momentum of the test particle $L$. In order to pursue the qualitative analysis of this function we have lowered the number of the parameters present in it, by introducing a new variable - $u=r/M$ and the parameters - related to the specific electrical charge of the black hole $b=(Q/M)^2$ and the quotient of angular momentum of the particle and mass of black hole $a=(L/M)^2$. We have found that the type of the trajectory described by a free test particle depends on these new parameters and not directly to the electrical charge $Q$, mass $M$ of the black hole and the angular momentum $L$ of the test particle. Moreover, the type of the trajectory depends on the total energy and initial position of the test particle.  

We have done the classification of the timelike geodesics of a GMGHS black hole separately for an arbitrary $b \in [0,2)$ and for $b=2$, because we have noticed that between the graphs of the corresponding effective potentials are differences, that may be reflected in the geometry of the orbits. Additionally, in order to visualize the trajectories described by test particles, we have determined numerical solutions of geodesic equations for different sets of the parameters. The values of the parameters were chosen using the knowledge accumulated during the qualitative study of the effective potential. The numerical solutions were plotted in polar coordinates. Our results are in good agreement with those obtained by Olivares and Villanueva (2013).

\textbf{Acknowledgments} The author would like to thank the anonymous reviewer for suggestions.

\end{document}